\documentstyle[preprint,version2,aps]{revtex}
\begin{document}
\preprint{Phys. Rev. Lett., 7 June 1993}
\draft

\begin{title}
Electron--Electron Scattering in Quantum Wires\\
and it's Possible Suppression due to Spin Effects
\end{title}

\author{Gerhard Fasol and Hiroyuki Sakaki}

\begin{instit}
Research Center for Advanced Science and Technology (RCAST),\\
University of Tokyo, 4--6--1 Komaba, Meguro--ku, Tokyo 153,
Japan
\end{instit}

\begin{abstract}

A microscopic picture of electron-electron pair scattering in single mode
quantum wires is introduced which includes electron spin. A new source of
`excess' noise for hot carriers is presented. We show that zero magnetic
field `spin' splitting in quantum wires can lead to a dramatic
`spin'-subband dependence of electron--electron scattering, including the
possibility of strong suppression. As a consequence extremely long electron
coherence lengths and new spin-related phenomena are predicted. Since
electron bands in III-V semiconductor quantum wires are in general
spin-split in zero applied magnetic field, these new transport effects are
of general importance.

\end{abstract}


\narrowtext

We show  that electron-electron pair scattering in quantum wires is
fundamentally different from two-dimensional (2D) or three-dimensional (3D)
systems and that it is essential to include the electron spin in the
analysis. We show that `spin' splitting of the electron bands causes
`spin'-subband dependence of electron pair scattering rates, and may cause a
dramatic reduction of electron-electron scattering for hot electrons in one
of the two spin subbands. We show that electron pair scattering can cause
fluctuations of electron spin, energy and wave number, and therefore is
expected to contribute `excess' noise to electrical current.  We expect that
spin related effects in quantum wire transport, as demonstrated in the
present work, will become important in mesoscopic transport experiments.

Electron-electron pair scattering is for many conditions the strongest
scattering mechanism, limiting the electron life-time and the phase coherence
length.  More generally, electron-electron interactions cause or contribute
to such diverse phenomena as superconductivity, Wigner crystallization,
magnetic ordering and heavy Fermion effects. In two dimensional (2D) and
three dimensional (3D) systems (but not in 1D) electron  pair scattering
contributes to thermalization of hot electrons. Although it does not
contribute directly to diffusive transport, it enters the collision integral
of the Boltzmann equation and thus in 3D and 2D (not in 1D) contributes to
establish diffusive transport. The general properties of electron pair
scattering in 3D have been investigated extensively \cite{pines}, but
detailed quantitative information for experimental semiconductor structures
has only become available recently in 2D \cite{yacoby}, \cite{fasolapl2D},
\cite{fasolssc}, and for 1D \cite{fasolapl1D}.  The  weak localization
regime, where impurity scattering dominates, has been intensively
investigated, but little is known about the properties of electron-electron
scattering in high mobility quantum wires. We assume in the present work,
that electrons in a quantum wire form an ordinary Landau liquid as supported
by  Ref.~\cite{dassarma} and that disorder effects are negligible. The
present work concerns single mode quantum wires, with a single (or a few)
transverse modes per spin orientation, i.e. wires with widths of the order of
100\AA .

First, we show that electron-electron pair scattering is a phase breaking
scattering process even in a `single-mode' quantum wire. Fig.~\ref{fig1}a
demonstrates such a process:  a `spin-up' electron at $\left( {\bf p},
\uparrow \right)$  scatters with a `spin-down' electron at $\left( {\bf k},
\downarrow \right)$,  resulting in a hole at $\left( {\bf k}, \downarrow
\right)$,  an electron at $\left( {\bf k}, \uparrow \right)$ and an electron
at $\left( {\bf p}, \downarrow \right)$.  Electron pair scattering in a
single mode quantum wire can only occur for pairs of electrons in opposite
`spin'-subbands, while it is forbidden for pairs of electrons in the same
`spin'-subband due to energy and momentum conservation and the Pauli
principle. Similarly, it was shown in  Ref.~\cite{fasolssc}, that in a 2DEG
scattering for pairs of electrons in the same `spin'-subband is typically
50\%  weaker than for opposite `spin'-subbands, although not totally
forbidden as in 1D.

Fig.~\ref{fig1}b shows the resulting picture for a non-equilibrium electron
propagating in a quantum wire. Electron pair scattering flips electrons
between the two different `spin'-subbands. Therefore, pair scattering leads
to fluctuations of electron spin. In general, the two `spin'-subbands in a
quantum wire will be split in energy, and the spin states will be mixed. As
Fig.~\ref{fig1}b demonstrates,  pair scattering in the presence of `spin'
subband splitting causes  fluctuations of electron energy and wave number in
addition to the spin fluctuations of a propagating electron, and therefore
should be experimentally important in quantum wire devices as a new
contribution to current dependent `excess'  noise.

In bulk III-V semiconductors, bands are spin split  at zero applied magnetic
field in all directions except $\left[ 100 \right] $  due to the lack of
inversion symmetry (see \cite{cardona}). Spin splitting has terms
proportional to $k$ and $k^{3}$, typical bulk values are shown in the insert
of Fig.~\ref{fig2}. The equivalent magnetic fields which would have to be
applied externally to produce a similar splitting at a Fermi energy around
$20meV$ are quite large. In quantum wells and quantum wires, terms in
addition to the bulk terms are expected \cite{roessler} \cite{bastard}.
Spin-splittings for 2D systems have recently been measured  \cite{jusserand}
\cite{das} \cite{dresselhaus}. For the rest of this work, we will show
results taking the conduction band structure equal to that of bulk GaAs. We
keep in mind that the precise value of the splitting and the spin mixing will
vary for different types of quantum wires, although there are always two
`spin' subbands in a `single' mode wire. We will not discuss sample dependent
details further in the present Letter, and we will simply label the two
subbands as `spin-up' and `spin-down'.

The essence of our results  can be explained with Fig.~\ref{fig2}. We consider
a pair scattering process, where an electron in the `spin-up' subband at
$\left( {\bf p}, \uparrow \right)$ scatters with a `spin-down' electron at
$\left( {\bf k}, \downarrow \right)$. Once ${\bf k}$ and ${\bf p}$ are
selected, the final states $\left({\bf k-q},\downarrow \right)$ and $\left(
{\bf p+q}, \uparrow \right)$ are determined by energy and momentum
conservation. (In the absence of spin splitting, or when the subbands are
parallel: ${\bf k - q = p}$). The probability for this process is given by
the product of the square of the Coulomb matrix element multiplied by the
thermal factor $f_{k,\downarrow }\left( 1 - f_{k - q, \downarrow } \right)
\left( 1 - f_{p + q,\uparrow }\right)$, where $f_{k}$\ldots\  are Fermi-Dirac
occupation factors. Clearly, spin-splitting strongly reduces the thermal
occupation probability factor for this scattering process. As a consequence,
forward ($k$ near $+k_{F}$) scattering is strongly suppressed for one
particular spin orientation (here `spin-up'), while there is a small increase
for the other spin orientation (here labelled `spin-down'). The strong `spin'
subband dependence of the scattering probability relies on the strong
k-dependence of the spin-splitting (bulk terms are proportional to $k$ and
$k^{3}$). It can be easily seen that `spin' subband dependent scattering
rates are not expected for k-independent splittings. Furthermore, for
scattering processes with $k \approx - k_{F}$ and $q \approx -2 k_{F}$  pair
scattering rates are almost independent of the subband. These facts weaken
the `spin' subband dependence of the total pair scattering rates, but
detailed calculations outlined below show, that in many circumstances strong
`spin' subband dependence prevails.

To confirm this surprising result quantitatively, we calculate the scattering
rates. The total scattering rate for an electron at wavevector $p, \sigma $ is
expressed as:

\begin{eqnarray}
\left. \frac {1}{\tau _{ee}} \right|_{p, \sigma } =
\frac {2 \pi }{\hbar}
\sum_ {k, q}
f_{k,\sigma' }\left( 1 - f_{k - q, \sigma' } \right)
\left( 1 - f_{p + q,\sigma }\right)
{\left| \frac
{\langle k-q, \sigma'; p+q, \sigma \left| V \right| k, \sigma';p, \sigma
\rangle }  {\epsilon \left( q, \left( E_{p, \sigma } - E_{p+q, \sigma }
\right) /\hbar \right) } \right|} ^{2} \nonumber \\
\times \
\delta \left( E_{p+q,\sigma } + E_{k-q, \sigma'} -
E_{p, \sigma } - E_{k, \sigma' } \right)
\label{eq2}
\end{eqnarray}

where  $\langle k-q, \sigma';p+q, \sigma \left| V \right| k, \sigma' ;p,
\sigma \rangle = e^{2} F^{1D}_{ijkl}(q \times w)/(L
\epsilon_{0}\epsilon_{r})$ is the 1D  Coulomb interaction matrix element.
$F^{1D}_{ijkl}(q \times w)$ is the 1D Coulomb Formfactor consisting of a
four-dimensional integral involving the wave functions and the Bessel
function $K_{0}$, which we determine by numerical integration assuming a
wire with a square cross section. The dielectric function $\epsilon \left( q,
\left( E_{p, \sigma } - E_{p+q, \sigma } \right)/\hbar \right)$ takes account
of dynamic screening. For the present calculation we integrate the finite
temperature Ehrenreich  expression for the polarizability numerically for the
two spin-split conduction bands. We assume that the quantum wire electron
band structure is described by the bulk k.p dispersion. The integrals are
calculated numerically using adaptive multipoint Gauss-Kronrod integration.

The details of an experimental quantum wire  will affect the band dispersion,
spin composition of the bands, the dielectric function and the matrix
elements. The essential point of the present letter is the prediction of a
large difference in the electron scattering rates for the two
`spin'-subband. The effects discussed in the present letter are a consequence
of the band splitting, the Pauli principle, Fermi occupation factors, and
energy and momentum conservation for electron pair scattering. They are
expected for many variations of the band structure, spin mixing and details
of the wave functions in different types of wires.

Fig.~\ref{fig3} compares the excess energy and spin-subband dependence of
differential pair scattering rates in a GaAs quantum wire for electrons in
the `spin--up' and `spin--down' subbands for a wire assumed to have the
conduction band structure of GaAs along $\left[ 110 \right]$. The carrier
concentration is $1.6 \times 10^{6} cm^{-1}$, temperature $T = 1.4K$,  and we
assume a square wire of width 100\AA\ and infinite confinement potential. Due
to the exponential character of the Fermi population factors, the  forward
pair scattering  rates for the `spin-up' subband are many orders of magnitude
lower compared to the `spin-down' subband. As expected, Fig.~\ref{fig3}b
shows that the `spin' subband dependence does not occur for electrons
scattering with partners at $k \approx - k_{F}$. Figures~\ref{fig3}a and b
clearly show, that for electrons with excess energies more than $1 meV$, the
total scattering rate is substantially larger for a hot electron in the
`spin-down' subband. We have investigated many combinations of `spin'
splitting strength, temperature and excess energy, and details will be
published separately. Constructing quantum wires with specific
`spin'-splitting, carrier concentration, and chosing particular temperature
and excess energy will allow to tune the spin-dependence of the electron pair
scattering rates.  Further it can be seen from Fig.~\ref{fig3}, that the
total scattering rates  also show some  spin dependence for equilibrium
electrons near the Fermi level ($\Delta = 0$), although the `spin'-subband
dependence is not strong.

Fig.~\ref{fig4} shows the total  electron pair scattering rates calculated
by numerically integrating Eq.~(\ref{eq2}) over $- \infty < k < + \infty $,
and the corresponding scattering lengths. The `spin' dependence of the
forward  scattering ($ k \approx + k_{F}$) causes a strong `spin' subband
dependence of the total rates.  For Millikelvin temperatures  scattering
lengths in excess of millimeters are predicted for one of the two
`spin'-subbands, while  at Helium temperatures lengths around  $20 \mu m$ are
predicted. To observe such long scattering lengths, other competing
scattering mechanisms have to be sufficiently weak. Impurity scattering and
interface roughness scattering can be reduced by improvements in fabrication
techniques. In Ref.~\cite{fasolapl1D} it was estimated, that remote ionized
impurity scattering can also be reduced sufficiently. The dashed line in
Fig.~\ref{fig4} shows the strength of acoustic phonon scattering in 2D from
Ref.~\cite{stormer}, comparable data for 1D are not yet available. Stronger
spin splitting due to choice of a different material, or in-built electric
fields, may lead to stronger suppression of scattering and longer coherence
lengths. Fig.~\ref{fig4} also shows that the `spin'-subband dependence of the
scattering rates disappears above a temperature larger than the typical
splitting energy of here $0.85 meV$, indicated in Fig.~\ref{fig2}, although
this temperature may be much increased for materials with higher spin
splitting.

We will now comment on the significance and on experimental predictions. We
have introduced a microscopic picture for electron pair scattering in single
mode quantum wires and calculated scattering rates. Such work is essential to
understand microscopic details of transport, or other details such as
spin-relaxation, which has recently attracted attention in 2D \cite{damen}.
We also demonstrated a new source of `excess' (i.e. current induced) noise.
The predicted `spin'-subband dependence of electron pair scattering rates
leads to the prediction of a range of novel `spin'-subband dependent
transport properties. The present work demonstrates that electron spin can
have even more dramatic effects in quantum wires than in 2D. Investigations
of mesoscopic transport have progressed to the point where very detailed
electronic spectroscopy of quantum dots coupled to quantum wire electron wave
guides can be performed (see e.g. Ref.~\cite{johnson}). `Spin' subband
dependence may allow high resolution experiments of magnetic sublevels in
quantum dots, and it may lead to electron spin polarization effects in hot
electron transport.

In summary, we have introduced a microscopic picture of electron
pair scattering for  quantum wires, which includes spin. We demonstrated it to
be a source of phase breaking and  `excess' noise. We show that `spin'
splitting leads to unequal forward and total pair scattering rates for
electrons in the `spin--up' and `spin--down' subbands of a quantum wire. We
predict the possibility of strong reduction of pair scattering for one of the
two `spin'-subbands, and very long `spin' subband dependent coherence
lengths. Several other spin-related effects may arise as a consequence of
`spin'-subband dependent electron pair scattering rates.

\acknowledgments

The authors would like to express their gratitude to Professor Yasuhiko
Arakawa,  Dr. Yasushi Nagamune and Dr. Jun-Ichi Motohisa for helpful
discussions. Support of this work by the NTT Endowed Chair at RCAST and the
`Sakaki Quantum Wave Project'  of ERATO is gratefully acknowledged. GF would
like to express his gratitude to Professor Manuel Cardona for introducing him
to  spin splitting and to Dr. Bernard Jusserand for stimulating discussions
and for sending preprints prior to publication.

\narrowtext

\figure{Microscopic picture of typical electron-electron pair scattering
process in a single mode quantum wire. {\bf (a)}~`Spin-up' electron at
$\left( {\bf p}, \uparrow \right)$ scatters with `spin-down' electron at
$\left( {\bf k}, \downarrow \right)$. Pair scattering of electrons in the same
`spin' subband is forbidden in 1D. {\bf (b)}~Electron pair scattering in a
quantum wire causes fluctuations of the electron spin. If the bands are
`spin'-split, electron energy and wave vector fluctuate as well, giving rise
to a new source of `excess' noise for a current of hot carriers.
\label{fig1}}

\narrowtext
\figure{Schematic diagram of electron--electron pair scattering process in a
quantum wire with spin splitting. An electron $\left( {\bf p}, \uparrow
\right)$ is scattered by electron $\left( {\bf k}, \downarrow
\right)$. Diagram shows typical spin--splitting of the conduction band
expected in a quantum wire along GaAs $\left[ 110 \right] $ near the Fermi
energy in zero applied magnetic field. Note that spin composition of the
bands is mixed and dependent on the details of the wire. For temperatures low
compared to the energy separation of states  $\left( {\bf p+q}, \uparrow
\right)$ and $\left( {\bf k}, \downarrow \right)$ (here approximately 0.85
meV as indicated),  the population factors entering the scattering
probability will lead to a dramatic  suppression of forward pair scattering
for electrons in the `spin-up' subband and to an enhancement of the
scattering rate for the `spin-down' subband. Insert shows typical values for
spin splitting in the bulk.  \label{fig2}}

\narrowtext
\figure{Differential scattering rates for  electrons in the `spin--up' and
`spin--down' subbands  of a quantum wire with `spin' splitting.
{\bf (a)}~As a consequence of the different Fermi population factors
electron--electron scattering with partners near $+ k_{F}$  is substantially
lower for one particular spin orientation (here spin up), while scattering for
the opposite spin orientation (here spin down) is enhanced. (Numerical
anomalies at $q=0$, where no dephasing takes place, are eliminated from the
figure). {\bf (b)}~For scattering with partners near $- k_{F}$ the scattering
rates are essentially independent of the `spin' subband.
\label{fig3}}

\narrowtext
\figure{Total electron--electron pair scattering rates determined by
numerical integration of Equ.~(\protect \ref{eq2}) over $-\infty < k <
+\infty $. Results are shown for a quantum wire, with conduction subband
dispersions assumed to be those of bulk GaAs oriented along the  $\left[ 100
\right]$ and $\left[ 110 \right]$ crystal orientations. Due to spin splitting
the total scattering rates are strongly suppressed for one of the two `spin'
subbands (here `spin--up'), while they are increased for the opposite `spin'
subband. Dashed line shows acoustic phonon scattering  for a 2DEG from
Ref.~\protect \cite{stormer}.
\label{fig4}}

\end{document}